# Static and Dynamic Routing, Fiber, Modulation Format, and Spectrum Allocation in Hybrid ULL Fiber-SSMF Elastic Optical Networks

**KANGAO OUYANG[1], FENGXIAN TANG[2], ZHILIN YUAN[1], JUN LI[1*] and YONGCHENG LI[1*]**
[1]School of Electronic and Information Engineering, Soochow University, Jiangsu Province 215006, P. R. China
[2]School of Integrated Circuits, Shenzhen Polytechnic University, Shenzhen, Guangdong Province, 518000, P. R. China

*Corresponding author: Yongcheng Li (ycli@suda.edu.cn) and Jun Li (ljun@suda.edu.cn)

This work was jointly supported by the supported by the National Key R&D Program of China (2023YFB2905505), the National Natural Science Foundation of China (NSFC) (62271338& 62201364), and the Natural Science Foundation of Jiangsu High Educational Institutions (22KJB510041).

**ABSTRACT** Traditional standard single-mode fibers (SSMF) are unable to satisfy the future long-distance and high-speed optical channel transmission requirement due to their relatively large signal losses. To address this issue, the ultra-low loss and large effective area (ULL) fibers are successfully manufactured and expected to deployed in the existing optical networks. For such ULL fiber deployment, network operators prefer adding ULL fibers to each link rather than replace existing SSMFs, resulting in a scenario where both of SSMF and ULL fiber coexist on the same link. In this paper, we investigated the routing, fiber, modulation format, and spectrum allocation (RFMSA) problem in the context of an elastic optical network (EON) where ULL fiber and SSMF coexisting on each link under both the static and dynamic traffic demands. We formulated this RFMSA problem as a node-arc based Mixed Integer Linear Programming (MILP) model and developed Spectrum Window Plane (SWP)-based heuristic algorithms based on different fiber selection strategies, including spectrum usage based (SU), optical signal-to-noise ratio (OSNR) aware, ULL fiber first (UFF), and random strategies. Simulation results show that in the static traffic demand situation, the RFMSA algorithm based on the OSNR-aware (OA) strategy exhibits optimal performance, attaining a performance similar to that of the MILP model regarding the maximum number of frequency slots (FSs) used in the entire network. Moreover, in the dynamic traffic demand scenario, the SU strategy remarkably surpasses the other strategies in terms of the lightpath blocking probability.

## I. INTRODUCTION

The explosive demand for network bandwidth been spurred by the novel Internet applications, such as cloud/edge computing, the Internet of Things (IoT), and augmented reality/virtual reality (AR/VR). This trend marks the advancement of optical networks that utilize standard single-mode fibers (SSMFs) as the major transmission medium, evolving from 100G/200G speeds to 400G and possibly 800G, thereby meeting the ever-increasing traffic demands [1]. However, as transmission rates increase, existing optical networks are approaching their capacity limits primarily due to the high fiber attenuation coefficient of SSMFs. This is because high-speed optical channels experience significant signal loss when traversing SSMFs, ultimately affecting the transmission distance and quality. To address this issue, ultra-low loss and large effective area (ULL) fibers have emerged as a promising new transmission medium for the next generation of optical networks, mainly due to their lower loss and larger effective area in comparison with SSMF [2][3]. For instance, network operators such as China Telecom, China Mobile, and China Unicom have each successfully completed 1000 km-level ultra-low-loss (ULL) fiber link transmission systems as cited in [4-6], thereby verifying the crucial role of ULL fiber in improving network transmission capacity. The large-scale deployment of ULL fibers in existing optical networks is bound to become a trend. There exist two approaches for deploying ULL fibers into existing optical networks. One is to replace all SSMFs and the other is to add ULL fibers alongside them. Network operators are more inclined to adopt the second approach since the existing SSMFs are still within their lifecycle [7]. By adopting the second approach for ULL fiber deployment, there will ultimately be one original SSMF and one new ULL fiber on each link in the network, which is called a hybrid ULL fiber-SSMF (HUS) optical network. It should be noted that in such a network, any lightpath is allowed to pass through different types of fibers, which is distinct from that of the conventional multi-fiber optical networks (MFON). This will give rise to a crucial issue of how to select the appropriate fiber on each link along a lightpath for the HUS optical network.

In [8-10], we have conducted research regarding the issue of optimal ULL fiber deployment in elastic optical networks, where routing, modulation format, and spectrum allocation (RMSA) algorithm are also deployed to solve the lightpath provisioning issue in the HUS optical network. However, [9] and [10] only consider the case where there is a single fiber on each link, while [8] considers the case where SSMF and ULL fiber coexist on a link. In [8], we developed a slot-window-plane (SWP) -based algorithm for provisioning lightpaths in the hybrid ULL fiber-SSMF elastic optical networks (HUS-EONs). However, this algorithm requires comparing the ULL fiber with the SSMF fiber on each link along the lightpath in every attempt and then making a selection, which has a relatively high computational complexity. When the network scale is relatively large, it is challenging to obtain results within a short time to support the online scheduling of traffic demands. Additionally, this study mainly focuses on the static traffic demands and fails to propose a corresponding algorithm for the dynamic traffic demands.

To solve these problems, we conducted a more in-depth study on the routing, fiber, modulation format, and spectrum allocation (RFMSA) problem of HUS-EON, and addressed it under both the static and dynamic traffic demands in this paper, respectively. The key contributions are as follows. We formulate the RFMSA problem in the context of HUS-EONs as a node-arc Mixed-Integer Linear Programming (MILP) model. Additionally, we propose efficient RFMSA algorithms based on different fiber selection strategies, such as Random, ULL fiber first (UFF), OSNR aware (OA), and spectrum usage (SU) based strategies. Note that, the first two strategies are applicable to both static and dynamic traffic demand scenarios, while the third strategy is only applicable to static traffic demand scenarios, and the last one is solely applicable to dynamic traffic demand scenarios. Simulation results indicate that in the static traffic demand scenario, the RFMSA algorithm based on the OA strategy attains the lowest maximum number of frequency slots (FSs) used and shows performance comparable to that of the MILP model. Furthermore, in the dynamic traffic demand scenario, the SU strategy remarkably outperforms the other strategies when considering lightpath blocking probability.

The rest of this paper is structured as follows: In Section II, we review related work on static and dynamic RMSA problem in EONs with multi-fiber or multi-core fiber on each link. Section III introduces the problem statement of lightpath provisioning in the HUS-EON. Section IV formulates this problem into the MILP model. Section V presents heuristic RFMSA algorithms and four different fiber selection strategies under static and dynamic traffic demand scenarios in HUS-EON. Section VI introduces simulation and performance analysis for different scenarios. Finally, we conclude the entire paper in Section VII.

## II. RELATED WORK

Currently, there are relatively few studies on the RFMSA problem for multi-fiber EONs. In contrast, a large number of studies have been carried out on routing, core, and spectrum assignment (RCSA) for the space-division multiplexing EON (SDM-EON) based on multi-core fibers. Since fiber allocation is relatively similar to core allocation, they will be introduced together in the following content.

For the static traffic demand scenario, Wu *et al.* proposed a new cost function that is pluggable into an auxiliary layered-graph framework to solve the routing, fiber, waveband, and spectrum assignment (RFBSA) problem with different objectives in multi-fiber EONs [11]. Yang *et al.* defined this lightpath provisioning problem in SDM-EONs as the routing, spatial channel, and spectrum assignment (RSCSA) problem and proposed an ILP model and a heuristic algorithm to solve the RSCSA problem [12]. Mrad *et al.* also developed a heuristic algorithm for addressing the RSCA problem in the SDM-EON [13]. Tang *et al.* tackled the routing, spectrum, core, and time assignment (RSCTA) problem for the SDM-EONs by developing an ILP model, as well as an auxiliary graph (AG) based heuristic algorithm [14]. Other studies on RCSA also include references [15-17].

For the dynamic traffic demand scenario, Wu *et al.* proposed a routing and spectrum assignment scheme that optimizes the state of the multi-fiber EONs and indicated its effectiveness by simulations [18]. Hirota *et al.* also proposed a dynamic RSA algorithm exploiting multiple fibers for multi-domain EONs, which enhances the statistical multiplexing effects with multiple fibers and reduces blocking probability [19]. Sharma *et al.* proposed a fragment-aware RCMSA algorithm for SDM-EON to mitigate the effects of fragments and reduce the blocking ratio while maintaining damage from crosstalk within threshold limits [20]. Yang *et al.* presented a RCSA scheme by preferentially locating the core with higher transmission quality, significantly reducing blocking probability and the resource fragmentation for the SDM-EON [21]. Chen *et al.* proposed a RMCSA algorithm that synthetically considers crosstalk and fragmentation for the SDM-EON, which provides a lower blocking probability and greater spectrum utilization [22]. Trindade *et al.* developed a fragmentation-aware RMSA algorithm in SDM-EONs, showing effectiveness in reducing significantly the blocking probability while respecting the inter-core cross-talk constraint in multi-core fiber [23].

According to our literature review, most studies mainly concentrate on multi-fiber networks based on SSMFs and do not take into account the coexistence of SSMFs and ULL fibers. In this study, we tackle the RSA problem in the HUS-EON and formulate it as a node-arc based mixed integer linear programming (MILP) optimization model. Four different fiber selection strategies and corresponding efficient heuristic algorithms are also proposed under both static and dynamic traffic demand scenarios.

### III. ROUTING, FIBER, MODULATION FORMAT, AND SPECTRUM ALLOCATION IN A HUS-EON

In this section, we first introduce the problem of lightpath provisioning in a HUS-EON. Then, we present the MILP formulation for the RFMSA problem for the HUS-EON under the static traffic demand scenario.

#### A. LIGHTPATH PROVISIONING IN A HUS-EON

Fig. 1 illustrates an example of provisioning lightpaths in a four-node and four-link HUS-EON under different fiber selection scenarios, including the case where the lightpath only occupies the spectrums of SSMF on each link, the case where the lightpath only occupies the spectrums of ULL fiber on each link, and the case where the lightpath occupies both SSMF and ULL. We consider three traffic demands within the network, namely A-B, B-C, and C-D, each requiring bandwidth of 160 Gb/s, 170 Gb/s, and 180 Gb/s, respectively. The candidate routes for each traffic demand are A-B, B-C-D, and C-D, respectively. For each lightpath, the corresponding modulation format is selected and the number of FSs is allocated in accordance with the OSNR of its candidate route. Note that the OSNR of each candidate can be calculated according to the formulas in Ref. [24] and the required FSs number of each traffic demand is calculated as the following formula.

$$F = \lceil B/E \rceil \quad (1)$$

where $B$ represents the bandwidth required by the traffic demand, and $E$ corresponds to the FS capacity associated with the modulation format selected, as shown in Table I.

Fig. 1(a) shows the scenario where lightpaths always select the SSMF on each link for establishment. Specifically, as the selected fibers, the OSNRs of different lightpaths are calculated as 14 dB, 10 dB and 13 dB respectively. Then, based on the OSNR, the corresponding modulation formats for the lightpaths A-B, B-C, and C-D are QPSK, BPSK, and QPSK, and their numbers of FSs can be determined as 4, 7 and 11, respectively. Thereby, the maximum number of FSs used in the whole network is 11. Note that, the maximum number of FSs used is defined as the largest index of FSs used in any fiber (regardless it is SSMF or ULL fiber) in the network.

Fig. 1(b) depicts the scenario where lightpaths consistently select the ULL fiber on each link for establishment. Due to the enhanced transmission performance provided by the ULL fiber, the candidate routes of the lightpaths possess higher OSNRs, which are 19 dB, 16 dB, and 17 dB respectively. Consequently, the lightpaths A-B, B-C, and C-D can use more advanced modulation formats, i.e., 16-QAM, 8-QAM, and 8-QAM, respectively. According to the FS calculation formula, the FSs required for the lightpaths A-B, B-C, and C-D are 2, 3, and 6 respectively, and the maximum number of FSs utilized in the entire network is reduced to 6.

In Fig. 1(c), we adaptively select appropriate fibers for the links along the candidate route of the lightpath. As shown in the figure, the lightpaths A-B and C-D still select SSMF on the links traversed by their candidate routes, while the lightpath B-C-D selects the SSMF on link B-C and the ULL fiber on link C-D. The OSNRs of the different lightpaths are 14 dB, 14 dB, and 13 dB, respectively, and the corresponding modulation formats are QPSK, QPSK, and QPSK, respectively. The required number of FSs of the lightpaths A-B and C-D remains constant, and that of the lightpath B-C-D is reduced to 4. Eventually, the maximum number of FSs used in the entire network is reduced to 4 because both the FSs on SSMF and ULL are used on link C-D. The example clearly shows that the selection of fibers in the links has a significant impact on the utilization of network spectrum resources. Therefore, it is crucial to investigate the RFMSA problem of HUS-EONs for achieving high spectrum efficiency.

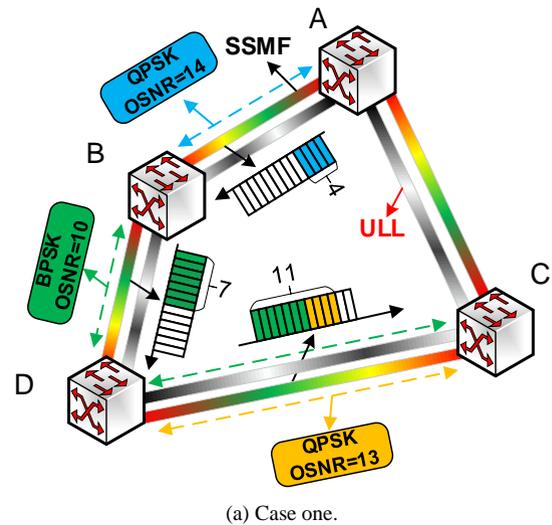

(a) Case one.

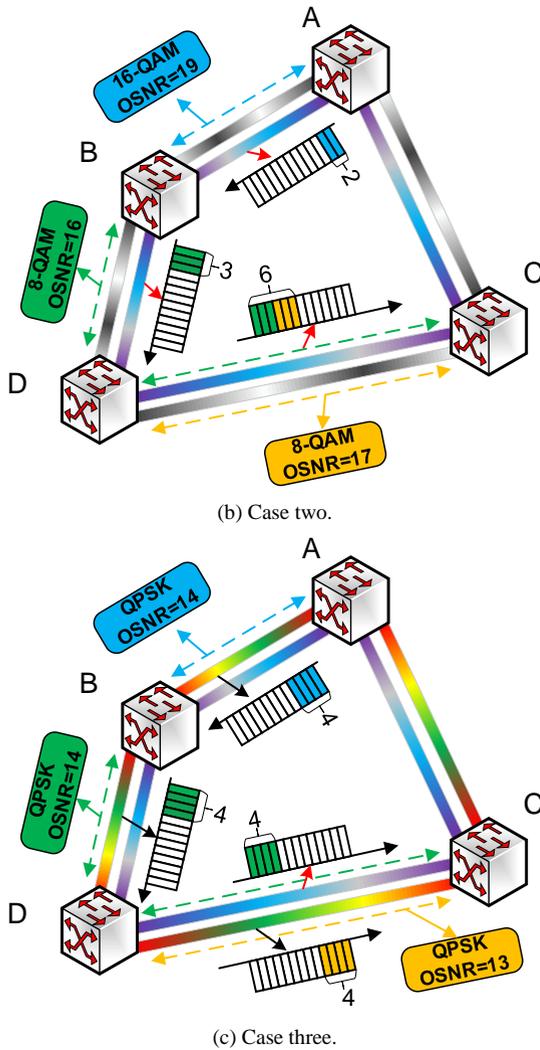

(b) Case two.

(c) Case three.

**FIGURE 1.** An example of provisioning lightpaths in a HUS-EON under three fiber selection scenarios.

### B. PROBLEM STATEMENT
Next, we define the RFMSA problem in the HUS-EON under the static traffic demand scenario as follows.

**Given:**
1) A general network topology represented by a graph $G_p(N, L)$, where $N$ is a set of nodes and $L$ is a set of links;
2) A set of static traffic demands given a priori, each demand is represented by a tuple $R(s, d, \lambda_{sd})$, where $s$ and $d$ are the source and destination nodes of the traffic demand, $\lambda_{sd}$ measured in units of Gb/s, represents the amount of traffic demand between node pair $(s, d)$.

**Constraints:**
1) All static traffic demands must successfully establish lightpaths;
2) All established lightpaths must satisfy spectrum non-overlapping, spectrum continuity, and spectrum contiguity;
3) The OSNR of all lightpaths must not be lower than the OSNR threshold of their selected modulation format;
4) Only one fiber can be chosen and it must be selected when selecting the fiber in the link along the route;

**Objective:** Minimize the maximum number of FSs used in the entire network.

### IV. MILP MODE
In this section, we formulate the RFMSA problem as a node-arc based MILP model for the HUS-EON under the static traffic demand scenario. The objective is to minimize the maximum number of FS used in the network. The sets, parameters, and variables of the model are as follows.

**Sets:**
$L$      Set of links. Link $ij \in L$ means that its two end nodes are $i$ and $j$.
$N$      Set of nodes.
$R$      Set of node traffic demands. Traffic demand $sd \in R$ means that its source and destination nodes are $s$ and $d$.
$P_n$      Set of links, each of which starts or ends at node $n \in N$.
$Q_{sd}$      Set of starting nodes of traffic demand $sd \in R$.
$E_{sd}$      Set of ending nodes of traffic demand $sd \in R$.
$M$      Set of candidate modulation formats for lightpath establishment, including BPSK, QPSK, 8-QAM, 16-QAM, 32-QAM, and 64-QAM.

**Parameters:**
$\beta$      Total number of FSs available in both the ULL fiber and SSMF.
$f_m^{sd}$      Number of FSs required for establishing a lightpath for traffic demand $sd$ when modulation format $m$ is used.
$OSNR_{ij,rec}^U$      A linear value that equals the reciprocal of lightpath OSNR contributed by ULL fiber on link $ij$.
$OSNR_{ij,rec}^S$      A linear value that equals the reciprocal of lightpath OSNR contributed by SSMF fiber on link $ij$.
$OSNR_{rec}^m$      A linear value that equals the reciprocal of the OSNR threshold required for establishing a lightpath using modulation format $m$.
$\nabla$      A large value.

**Variables:**
$\gamma_{ij}^{sd}$      A binary variable that equals 1 if the route of traffic demand $sd$ traverses link $ij$; 0, otherwise.
$\rho_{sd}^n$      A binary variable that equals 1 if the lightpath of traffic demand $sd$ traverses node $n$; 0, otherwise.
$S_{sd}$      An integer variable, indicating the starting FS index of the lightpath of traffic demand $sd$.
$X_{sd1,sd2}$      A binary variable that equals 1 if the lightpath starting FS index of traffic demand $sd1$ is larger than the lightpath starting FS index of traffic demand $sd2$, i.e., $S_{sd1} > S_{sd2}$; 0, otherwise.
$Z_{sd,ij}$      A binary variable. If a lightpath established for traffic demand $sd$ uses ULL fiber on link $ij$, it equals 1; 0, otherwise.
$E_{sd,ij}$      A binary variable. If a lightpath established for traffic demand $sd$ uses SSMF on link $ij$, it equals 1; 0, otherwise.

| | |
|---|---|
| $\theta_{sd1,sd2}$ | A binary variable. If there is a shared fiber link between the lightpaths for traffic demands $sd1$ and $sd2$, it equals 1; 0, otherwise. |
| $OSNR_{sd,rec}$ | The reciprocal of the OSNR of a lightpath for traffic demand $sd$. This is a linear value. |
| $O_{sd,ij}$ | The reciprocal of the OSNR of a lightpath for traffic demand $sd$. This is a linear value. |
| $\delta_{sd,m}$ | A binary variable. If modulation format $m$ is chosen for a lightpath for traffic demand $sd$, it equals 1; 0 otherwise. |
| $F_{sd}$ | An integer variable that equals the number of FSs allocated to a lightpath for traffic demand $sd$. |
| $C$ | The maximum index of FSs used in the entire network. |
| $\eta_{ij}$ | An integer variable that equals the maximum number of FSs used on link $ij$. |
| $\phi_{ij}$ | An integer variable that equals the maximum index of FSs used in the ULL fiber of link $ij$. |
| $\varphi_{ij}$ | An integer variable that equals the maximum index of FSs used in the SSMF of link $ij$. |
| $\sigma_{sd,ij}$ | An integer variable that equals the maximum index of FSs used by the route for traffic demand $sd$ in the ULL fiber of link $ij$. |
| $\tau_{sd,ij}$ | An integer variable that equals the maximum index of FSs used by the route for traffic demand $sd$ in the SSMF of link $ij$. |
| $\xi_{sd,ij}$ | An integer variable that equals the maximum index of FSs used by a lightpath for traffic demand $sd$ in the ULL fiber of link $ij$. |
| $\zeta_{sd,ij}$ | An integer variable that equals the maximum index of FSs used by a lightpath for traffic demand $sd$ in the SSMF of link $ij$. |

**Objective:** Minimize $C$ (2)

**Constraints:**

-Route allocation

$$\sum_{ij \in P_n} \gamma_{ij}^{sd} = 1 \quad \forall sd \in R, \forall n \in Q_{sd} \quad (3)$$

$$\sum_{ij \in P_n} \gamma_{ij}^{sd} = 1 \quad \forall sd \in R, \forall n \in E_{sd} \quad (4)$$

$$\sum_{ij \in P_n} \gamma_{ij}^{sd} = 2 \cdot \rho_{sd}^n \quad \forall sd \in R, \forall n \in N - (Q_{sd} \cup E_{sd}) \quad (5)$$

-Fiber allocation

$$Z_{sd,ij} \leq \gamma_{ij}^{sd} \quad \forall sd \in R, \forall ij \in L \quad (6)$$

$$E_{sd,ij} \leq \gamma_{ij}^{sd} \quad \forall sd \in R, \forall ij \in L \quad (7)$$

$$E_{sd,ij} \leq (1 - Z_{sd,ij}) \quad \forall sd \in R, \forall ij \in L \quad (8)$$

$$E_{sd,ij} \geq \gamma_{ij}^{sd} + (1 - Z_{sd,ij}) - 1 \quad \forall sd \in R, \forall ij \in L \quad (9)$$

-Modulation format allocation

$$\sum_{m \in M} \delta_{i,m} = 1 \quad \forall sd \in R \quad (10)$$

$$O_{sd,ij} \leq \nabla \cdot \gamma_{ij}^{sd} \quad \forall sd \in R, \forall ij \in L \quad (11)$$

$$O_{sd,ij} \leq Z_{sd,ij} \cdot OSNR_{ij,rec}^U + E_{sd,ij} \cdot OSNR_{ij,rec}^S \quad \forall sd \in R, \forall ij \in L \quad (12)$$

$$O_{sd,ij} \geq Z_{sd,ij} \cdot OSNR_{ij,rec}^U + E_{sd,ij} \cdot OSNR_{ij,rec}^S - \nabla \cdot (1 - \gamma_{ij}^{sd}) \quad \forall sd \in R, \forall ij \in L \quad (13)$$

$$OSNR_{sd,rec} = \sum_{ij \in L} O_{sd,ij} \quad \forall sd \in R \quad (14)$$

$$OSNR_{sd,rec} - OSNR_{rec}^m \leq \nabla \cdot (1 - \delta_{sd,m}) \quad \forall sd \in R, \forall m \in M \quad (15)$$

-Spectrum allocation

$$F_{sd} = \sum_{m \in M} \delta_{sd,m} \cdot f_m^{sd} \quad \forall sd \in R \quad (16)$$

$$\theta_{sd1,sd2} \geq Z_{sd1,ij} + Z_{sd2,ij} - 1 \quad \forall sd1, sd2 \in R, \forall ij \in L: sd1 \neq sd2 \quad (17)$$

$$\theta_{sd1,sd2} \geq E_{sd,ij} + E_{sd,ij} - 1 \quad \forall sd1, sd2 \in R, \forall ij \in L: sd1 \neq sd2 \quad (18)$$

$$S_{sd2} - S_{sd1} \leq \nabla \cdot (1 - X_{sd1,sd2} + 1 - \theta_{sd1,sd2}) - 1 \quad \forall sd1, sd2 \in R: sd1 \neq sd2 \quad (19)$$

$$S_{sd1} + F_{sd1} - S_{sd2} \leq \nabla \cdot (X_{sd1,sd2} + 1 - \theta_{sd1,sd2}) \quad \forall sd1, sd2 \in R: sd1 \neq sd2 \quad (20)$$

-Maximum number of FSs used calculation

$$\xi_{sd,ij} \geq S_{sd} + F_{sd} - \beta \cdot (1 - Z_{sd,ij}) \quad \forall ij \in L, \forall sd \in R \quad (21)$$

$$\sigma_{sd,ij} \leq \nabla \cdot \gamma_{ij}^{sd} \quad \forall ij \in L, \forall sd \in R \quad (22)$$

$$\sigma_{sd,ij} \leq \xi_{sd,ij} \quad \forall ij \in L, \forall sd \in R \quad (23)$$

$$\sigma_{sd,ij} \geq \xi_{sd,ij} - \nabla \cdot (1 - \gamma_{ij}^{sd}) \quad \forall ij \in L, \forall sd \in R \quad (24)$$

$$\phi_{ij} \geq \sigma_{sd,ij} \quad \forall ij \in L, \forall sd \in R \quad (25)$$

$$\zeta_{sd,ij} \geq S_{sd} + F_{sd} - \beta \cdot (1 - E_{sd,ij}) \quad \forall ij \in L, \forall sd \in R \quad (26)$$

$$\tau_{sd,ij} \leq \nabla \cdot \gamma_{ij}^{sd} \quad \forall ij \in L, \forall sd \in R \quad (27)$$

$$\tau_{sd,ij} \leq \zeta_{sd,ij} \quad \forall ij \in L, \forall sd \in R \quad (28)$$

$$\tau_{sd,ij} \geq \zeta_{sd,ij} - \nabla \cdot (1 - \gamma_{ij}^{sd}) \quad \forall ij \in L, \forall sd \in R \quad (29)$$

$$\varphi_{ij} \geq \tau_{sd,ij} \quad \forall ij \in L, \forall sd \in R \quad (30)$$

$$\eta_{ij} \geq \phi_{ij} \quad \forall ij \in L \quad (31)$$

$$\eta_{ij} \geq \varphi_{ij} \quad \forall ij \in L \quad (32)$$

$$C \geq \eta_{ij} \quad \forall ij \in L \quad (33)$$

**Explanations of equations:**

**Route allocation**: Constraints (3)-(5) find a path for lightpath establishment between each traffic demand. Specifically, constraint (3) ensures that the first link of a lightpath for traffic demand $sd$ originates from node $s$. Constraint (4) ensures that the last link of a lightpath for traffic demand $sd$ ends at node $d$. Constraint (5) ensures that any intermediate node traversed by a lightpath for traffic demand $sd$ is associated with exactly two links.

**Fiber allocation:** Constraints (6)-(9) ensure that only one type of fiber (either SSMF or ULL fiber) in a link can be selected to establish a lightpath.

**Modulation format allocation**: Constraint (11)-(14) calculates the OSNR of a lightpath. Constraint (15) ensures that the OSNR of a lightpath meets the OSNR threshold required by a chosen modulation format.

**Spectrum allocation**: Constraint (16) calculates the number of FSs required to establish a lightpath. Constraints (17) and (18) check whether two different lightpaths share same fiber link. Constraints (19) and (20) ensure that the spectrum used by different lightpaths do not overlap on the shared fiber.

**Maximum number of FSs used calculation:** Constraints (21)-(25) find the maximum index of FSs used on each ULL fiber link. Constraints (26)-(30) find the maximum index of FSs used on each SSMF link. Constraints (31)-(33) find the maximum number of FSs used on these two types of fiber links.

The computational complexity of the MILP model is decided by the dominant numbers of variables and constraints. For the above MILP model, the dominant number of variables are decided the variables $X_{sd1,sd2}$ and $\theta_{sd1,sd2}$. Its number of variables is of order of $O(|R|^2)$, where $|R|$ is the total number of for traffic demands. Similarly, for the dominant number of constraints, we need to consider constraints (17)-(20), their dominant numbers of constraints are also of the order of $O(|R|^2)$.

## V. HEURISTIC ALGORITHMS

The MILP model is suitable for addressing optimization problems in small-scale networks. However, as the scale of the network increases, its computational complexity would be high, making it impractical to solve large-scale models within a reasonable amount of time. Therefore, in addition to the MILP model, we have also developed an efficient heuristic algorithm to address the optimization problem by extending the conventional SWP-based algorithm, called SWP-based RFMSA algorithm, in which we consider four different strategies to choose the fiber on each link along the lightpath, including Random, ULL fiber first (UFF), OSNR aware (OA), and spectrum usage (SU) strategies. Note that the OA strategy is only applicable to the static traffic demand scenario, while the SU strategy is only applicable to dynamic traffic demand scenario. Next, we will introduce the four fiber selection strategies first, followed by a detailed explanation of the SWP-based RFMSA algorithm.

### A. FIBER SELECTION STRATEGIES

We consider four strategies for choosing fiber on each link along the candidate of a lightpath. These strategies are employed in our proposed SWP-based RFMSA algorithm and are presented as follows:

**Random Strategy**: This strategy randomly selects either ULL fiber or SSMF for spectrum allocation on each link traversed by a lightpath. Specifically, in our proposed RFMSA algorithm (introduced later), when constructing a SWP within a given FSs range, if both ULL fiber and SSMF on a network link have these continuous FSs available, one fiber is randomly selected to be mapped as a virtual link to the SWP. If only one fiber has this spectrum available, that fiber is mapped to the SWP. If neither ULL fiber nor SSMF has the spectrum available within this range, the corresponding virtual link is unavailable on the SWP.

**UFF Strategy**: This strategy gives a priority to the utilization of ULL fiber for the lightpath establishment in the HUS-EON. Specifically, it initially maps the ULL fiber of each network link to each SWP under different modulation formats, and subsequently employs our proposed RFMSA algorithm to establish lightpaths for given traffic demands. In case all modulation formats have been attempted and the lightpath remains unestablished, the entire process is reiterated using SSMF in place of ULL fiber to construct SWPs.

**OA Strategy**: This strategy utilizes the OSNR values of ULL fiber and SSMF on each link for fiber selection. Specifically, when mapping a virtual link to an SWP, if only one fiber on the corresponding network link has the required contiguous FSs available, that fiber is directly mapped to the SWP. If both ULL fiber and SSMF are available, the selection is made by comparing their OSNR values. If the ratio of the OSNR of the ULL fiber to that of the SSMF exceeds **a predefined threshold $\alpha$**, the ULL fiber is mapped to the SWP; otherwise, the SSMF is mapped to the SWP.

**SU Strategy**: This strategy is only applicable to dynamic traffic demand scenario. Once an available route with $n$ links is found in a certain SWP, there will be $2^n$ different fiber selection schemes $\mathcal{H}$. This strategy will calculate the cost $c$ of each available fiber selection scheme according to the following formula and then select the one with the largest cost as the solution.

$$c_h = n_h^d \times \frac{B_h}{m} \times \Omega \quad h \in \mathcal{H} \qquad (34)$$

Here, $c_h$ is defined as the cost of the fiber selection scheme $h$. A fiber selection scheme with large value of $c$ indicates that it has more available spectrum capacity to serve the traffic demand. $n_h^d$ represents the total number of available spectrum blocks, i.e., contiguous idle FSs required by the traffic demand $d$, along the route when the fiber selection scheme $h$ is adopted. $B_h$ is defined as the total number of adjacent FS state changes along the route when the fiber selection scheme $h$ is employed, while $m$ represents the maximum number of adjacent FS state changes, which is fixed to $\beta - 1$. Here $\beta$ is the total number of FSs available in each fiber. $\Omega$ is a weight factor. If the fiber selection scheme $h$ always uses SSMFs along the route, $\Omega$ is set as 1. Otherwise, if the fiber selection scheme $h$ employs ULL fiber and can help reduce the number of FSs required for the traffic demand, it is set as 1.2. Conversely, it is set as 0.8.

Fig. 2 shows an example of choosing a fiber selection scheme for the lightpath A-B-C. The candidate route of the lightpath consists of two links, A-B and B-C. Thus, there are four different fiber selection schemes, $h1$, $h2$, $h3$, and $h4$, as shown in Fig. 2(b). $h1$ always selects SSMFs along the route A-B-C, while h2 selects ULL fibers in both links A-B and B-C. $h3$ selects ULL fiber in link A-B and SSMF in link B-C. $h4$ selects SSMF in link A-B and ULL fiber in link B-C. According to formula (1), we calculate the costs of each fiber selection scheme as 1.069, 3.05, 0.44, and 0.65. Finally, $h2$ is used to allocate fibers when establishing a lightpaths between the node pair (A, C).

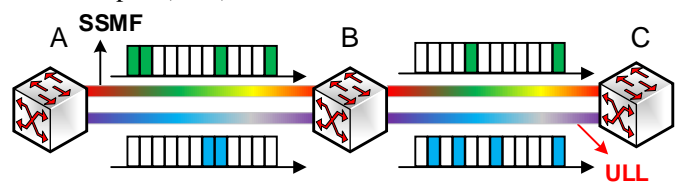

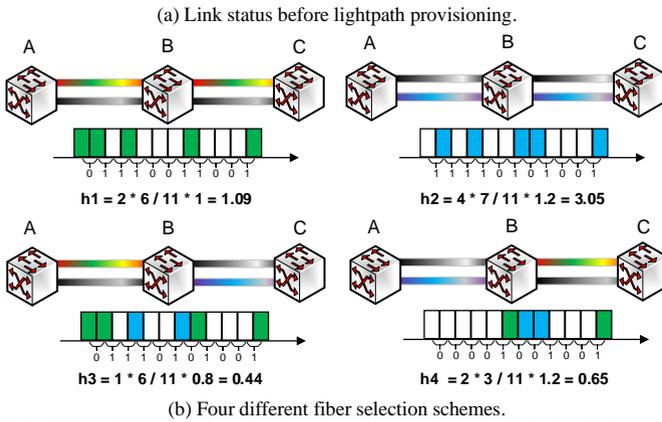

(a) Link status before lightpath provisioning.

h1 = 2 * 6 / 11 * 1 = 1.09
h2 = 4 * 7 / 11 * 1.2 = 3.05
h3 = 1 * 6 / 11 * 0.8 = 0.44
h4 = 2 * 3 / 11 * 1.2 = 0.65

(b) Four different fiber selection schemes.

**FIGURE 2.** An example of selecting fiber on each link for lightpath provisioning.

### B. SWP-BASED RFMSA ALGORITHM

We extend the conventional SWP-based heuristic algorithm [25] to solve the RFMSA problem, respectively, for efficiently provisioning lightpaths in a HUS-EON. The pseudocode is as follow.

| SWP-based RFMSA Algorithm |
|---|
| **Input:** a network topology $G_p = (N, L)$, a traffic demand $d$ between a node pair, an empty route $R_* = NULL$; |
| 1  **For** each $m \in M$ //$M$ is the set of available modulation formats sorting in descending order based on spectrum efficiency. |
| 2      Calculate the number of FSs required $f$ using formula (1) by $m$ and $d$; |
| 3      **If** $f == f'$ **then** // $f'$ is the number of FSs required by using the next modulation formulation $m'$ |
| 4          $m = m'$; $f = f'$; |
| 5      **End if** |
| 6      $\Theta = CreateSWPList(G_p, m, f)$; //Employ the strategies introduced in section V. A to generate SWP list |
| 7      **For** each $\omega \in \Theta$ |
| 8          Employ Dijkstra's algorithm to find the shortest route $R$ based on the physical distance of each virtual link in $SWP$; |
| 9          **If** SU strategy is employed **then** |
| 10             Generate the set of fiber selection schemes $\mathcal{H}$ based on the shortest route $R$. Remove those fiber selection schemes from $\mathcal{H}$ which cannot meet the threshold for the modulation format $m$. |
| 12            Calculate the cost of each fiber selection scheme in set $\mathcal{H}$, and find the one $h^*$ with the largest cost. |
| 13            $R_* = R$; |
| 14         **Else** |
| 15             **If** the OSNR of route $R$ is not less than the OSNR threshold of $m$ **then** |
| 16                 According to the different fiber selection strategies introduced in Section V.A, select an appropriate fiber on each link of the shortest route $R$; |
| 17                 $R_* = R$; |
| 18             **End If** |
| 19         **End If** |
| 20     **End for** |
| 21     **If** $R_* = NULL$ && $m$ is not the final element of $M$ **then** |
| 22         Return step 2; |
| 23     **Else** |
| 24         **If** $m$ is not the final element of $M$ **then** |
| 25             Block the request; |
| 26         **Else** |
| 27             Establish a lightpath along route $R_*$; |
| 28         **End If** |
| 29     **End If** |
| 30 **End for** |
| 31 **End** |

This algorithm tries to traverse all the modulation formats based on their spectrum efficiency from the highest to lowest. For each modulation format, first calculate the number of FSs required and compare it with that of next modulation format. If they are the same, directly traverse to the next modulation format.

In step 7, the algorithm generates a SWP list based on the current modulation format $m$. For each SWP, add virtual links corresponding to different fiber selection strategies mentioned in Section V.A. In steps 8, we employ Dijkstra's algorithm to find the shortest route $R$ based on the physical distance of each virtual link in the SWP.

In steps 9-20, this algorithm selects appropriate fibers on each link along the shortest route $R$. In steps 9-13, by employing the SU strategy, it selects the fiber selection scheme with the highest cost among all possible alternatives as the scheme for allocating corresponding fibers on each link of the shortest route R. In steps 15-17, the algorithm selects appropriate fibers for the links of the route according to the fiber selection strategies, Random, UFF, and OA, respectively.

In steps 21-30, algorithm tries to establish a lightpath for the traffic demand request. If $R_* == null$, the algorithm considers lowering the modulation format and repeats the process until all modulation formats have been attempted. Otherwise, a lightpath is established based on the found shortest route, corresponding SWP, and fiber selection scheme. If the lightpath still cannot be established after traversing all modulation formats, the traffic demand request will be blocked.

### VI. TEST CONDITIONS AND RESULT ANALYSES

In this section, we first introduce the test conditions of this study, and then perform performance analysis of different algorithms and fiber selection strategies for static and dynamic traffic demand scenarios, respectively.

## A. TEST CONDITIONS

To evaluate the performance of the proposed lightpath provisioning algorithms in a HUS-EON, we consider two test networks, i.e., the 6-node, 9-link (n6s9) network and the 24-node, 43-link (USNET) network, as shown in Fig. 3. Each link in the network consists of a SSMF and an ULL fiber. The ULL fiber is assumed to be the Corning's TXF fiber, with a typical attenuation coefficient of 0.166 dB/km [26], while the SSMF has an attenuation coefficient of 0.20 dB/km. There are 320 FSs in each fiber link, the bandwidth granularity of each FS is 12.5 GHz. Optical amplifiers (i.e., EDFAs) are placed at equal distances, no greater than 80 km.

We use six modulation formats (including BPSK, QPSK, 8-QAM, 16-QAM, 32-QAM, and 64-QAM) for lightpaths establishment. The FS capacity and the OSNR required by each modulation format are shown in Table I. Here, we consider two types of traffic demand scenarios, including static traffic demand and dynamic traffic demand, respectively. For the former, the traffic demand between each node pair is randomly generated with a uniform distribution ranging from [10, $X$] Gb/s, where $X$ is the maximum traffic demand between the node pair. For the latter, we assume that each node has a traffic load in Erlang units, where the arrival of lightpath requests follows a Poisson process, and their service holding times follow an exponential distribution. We utilized the commercial software AMPL/Gurobi (version 9.0.2) to solve the MILP model and used JAVA to implement the proposed heuristic algorithms.

TABLE I
FS CAPACITIES AND OSNR THRESHOLDS OF DIFFERENT MODULATION FORMATS [27]

| Modulation format | FS capacity (Gb/s) | OSNR Threshold (dB) |
|---|---|---|
| BPSK | 25 | 9 |
| QPSK | 50 | 12 |
| 8-QAM | 75 | 16 |
| 16-QAM | 700 | 18.6 |
| 32-QAM | 125 | 21.6 |
| 64-QAM | 150 | 24.6 |

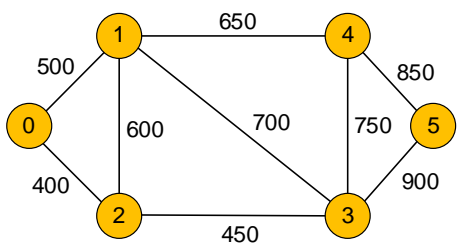

(a) 6-node, 9-link network (n6s9).

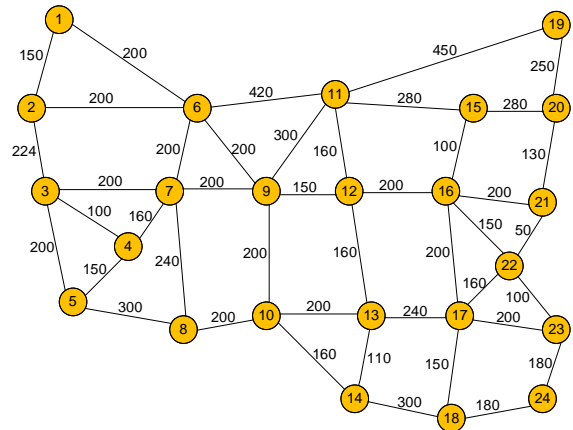

(b) 24-node, 43-link network (USNET).

**FIGURE 3.** Test networks.

## B. PERFORMANCE COMPARISON OF FIBER SELECTION STRATEGIES

We first evaluate the performance of different fiber selection strategies in terms of maximum number of FSs under static traffic demand and the lightpath blocking probability under the dynamic traffic demand, respectively. Here, the maximum number of FSs is defined as the largest index of FSs used in any fiber (regardless it is SSMF or ULL fiber) in the network and the lightpath blocking probability is defined as the ratio of the number of traffic demand requests that fail to establish the lightpath to the total number of traffic demand requests.

### 1) MAXIMUM NUMBER OF FSS USED

For the different strategies, we first compare the maximum number of FSs used. Fig. 4 compares the results of the n6s9 network, where $X$ is within the range of [15, 35] Gb/s. The legend "MILP" corresponds to the MILP model, "SWP" corresponds to the SWP-based RFMSA algorithm, "R" corresponds to the Random strategy, "UFF" corresponds to the UFF strategy, and "OA" corresponds to the OA strategy. Moreover, we also use the proposed RFMSA algorithm in reference [8] for comparison, which is represented as "Adaptive".

We can observe that the OA strategy demonstrates the best performance. This is attributed to the OA strategy's greater flexibility in fiber selection and can effectively utilize the ULL fibers. We also compare the performance of the heuristic algorithms and the MILP model. The result shows that the SWP-based RFMSA with OA strategy can achieve performance very close to the MILP model. This shows the effectiveness of the proposed heuristic algorithm in reducing the number of FSs used. Meanwhile, we can observe that the results of the OA strategy align with our previously proposed algorithm. However, the time complexity of the Adaptive strategy is $O(|M| \cdot |F|^2 \cdot |N|^2 \cdot |L| \cdot |T|)$, while the computation complexity of the OA strategy is only $O(|M| \cdot |F| \cdot |N|^2 \cdot |L|)$. Here, $|M|$ denotes the number of modulation formats, $|F|$ represents the available FSs in the network, $|N|$ and $|L|$ are the number of nodes and links in the network, and $|T|$ is the total number of hops of the route. This indicates that our proposed OA strategy can obtain results close to the original SWP-based RFMSA algorithm more quickly.

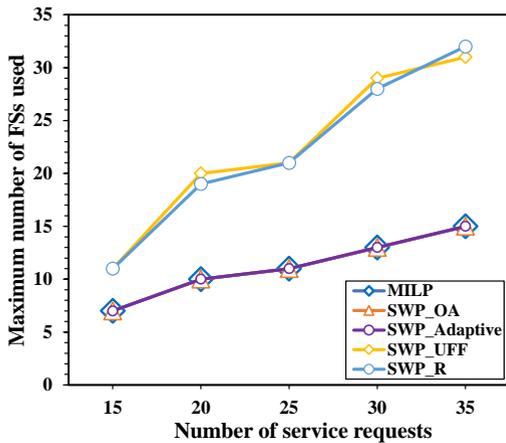

**FIGURE 4.** Maximum number of FSs used under different strategies in the n6s9 network.

Fig. 5 shows the results of USNET, in which a similar observation can be made for the four fiber selection strategies. Here, due to the high computational complexity of the MILP model, we do not provide its results. The $X$ is ranging from 100 to 700 Gb/s. Similar, we can note that the OA strategy demonstrates its effectiveness in reducing the maximum number of FSs used by up to 41.7%, and 26.2%, respectively, compared with the UFF and random strategy. This once again confirms the effectiveness of the OA strategy. Also, the results of the OA strategy are largely consistent with the algorithm we previously proposed when applied to large-scale networks. This further demonstrates the effectiveness of the OA strategy in terms of both time complexity and spectrum utilization performance.

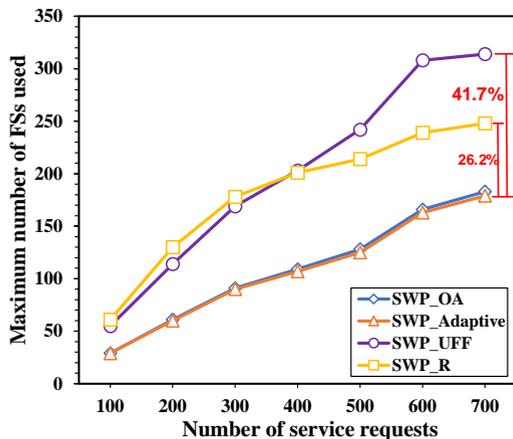

**FIGURE 5.** Maximum number of FSs used under different strategies in the USNET.

### 2) LIGHTPATH BLOCKING PROBABILITY

We also evaluate the blocking probability performance for different fiber selection strategies, where both the SWP-based and the shortest path-based RFMSA algorithms are employed. Fig. 6 shows the results of the n6s9 network and NSFNET network, respectively. The legends "SU" corresponds to the SU strategy, "SP" refers to the shortest path-based algorithm, and other legends have the same meaning as before.

Fig. 6(a) shows the results for the n6s9 network when the traffic load between each node pair increases from 30 Erlang to 42 Erlang. Comparing the different fiber selection strategies, we note that the SU strategy can significantly improve the lightpath blocking performance in the context of both RFMSA algorithms compared with the Random and UFF strategies. This is because the SU strategy selects fibers more flexibly according to the cost formula. It effectively reduces spectral fragmentation and consequently, lowers the probability of traffic demands being blocked.

In addition, we also note that the SWP-based RFMSA algorithm can perform significantly better than the shortest path-based RFMSA algorithm in the context of different fiber selection strategies. This is because the SWP-based RFMSA algorithm offers more flexible routing and spectrum allocation compared to the SP-based RFMSA algorithm, which in turn reduces the blocking probability.

Fig. 6(b) shows a similar performance comparison for USNET. We also note that the SU strategy can significantly improve the lightpath blocking performance for both the SWP-based and SP-based RFMSA algorithms compared with the Random and UFF strategies. Finally, the SWP-based RFMSA algorithm can significantly improve the lightpath blocking performance compared with the SP-based RFMSA algorithm.

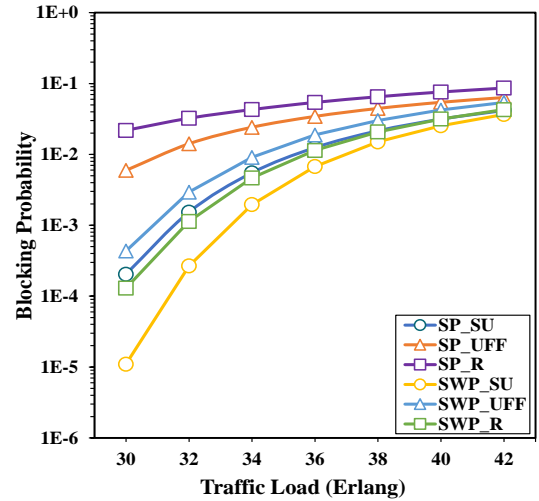

(a) n6s9 network.

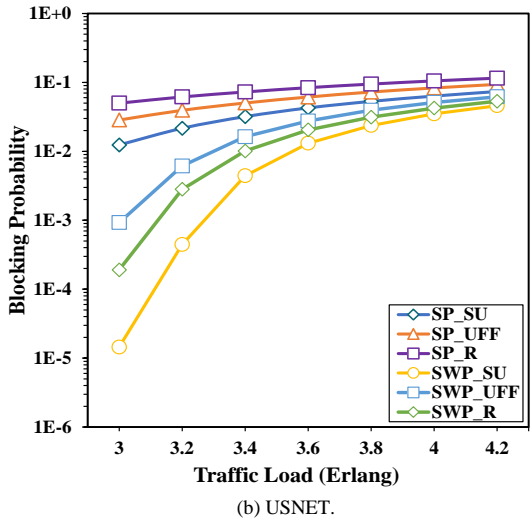

(b) USNET.
FIGURE 6. Blocking probability under different strategies.

### C. IMPACT OF LINK THRESHOLD α

The value of the threshold $\alpha$ may impact the performance of the OA strategy. To study this, we also evaluate the maximum number of FSs used of an HUS-EON using the SWP-based RFMSA algorithm with OA fiber selection strategy under different values of the threshold $\alpha$. Fig. 6 shows the results of the n6s9 network, where the value of threshold α increases from 1.1 to 1.14. The legends "200" and "300" represent the maximum traffic demand between the node pair $X$ are 200 Gb/s and 300 Gb/s, respectively.

It can be seen that the maximum number of FSs used decreases first and then tends to stabilize as the threshold α increases. Specifically, when the threshold reaches 1.12, continuing to increase the threshold will no longer reduce the maximum number of FSs used. This is because when the threshold α is low, the OA strategy tends to prioritize the use of ULL fiber for establishing lightpaths. This leads to a large number of FSs on ULL fibers being used, thus making the maximum number of used FSs in the entire network very large. As the threshold α increases, more SSMFs are used to establish lightpaths, thus balancing the number of FSs used on different fibers in each link and reducing the maximum number of FSs used in the network. However, with a further increase in α , since SSMFs cannot meet the OSNR requirements of traffic demands, no more SSMFs will be selected. Therefore, the maximum number of FSs used in the network will not further decrease. In addition, there is a slight increase of the maximum number of FSs used lastly because the selection of SSMFs will cause traffic demands to require more FSs, which may potentially increase the maximum number of FSs used in the entire network.

We have a similar conclusion for USNET network. Fig. 7(b) shows that with increasing threshold, the maximum number of FSs used in the network initially decreases gradually, then slightly increases, which is consistent with the results in the n6s9 network. The optimal network performance is achieved when the link threshold is set to 1.09.

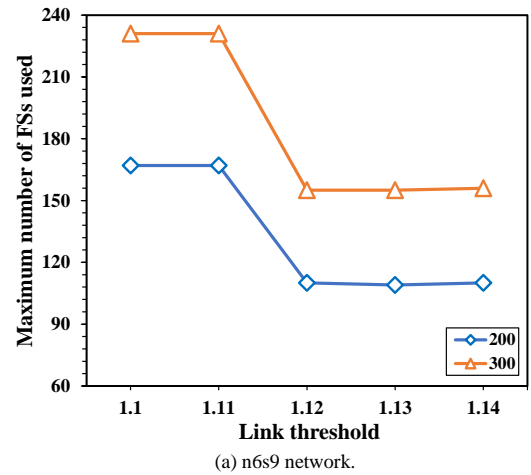

(a) n6s9 network.

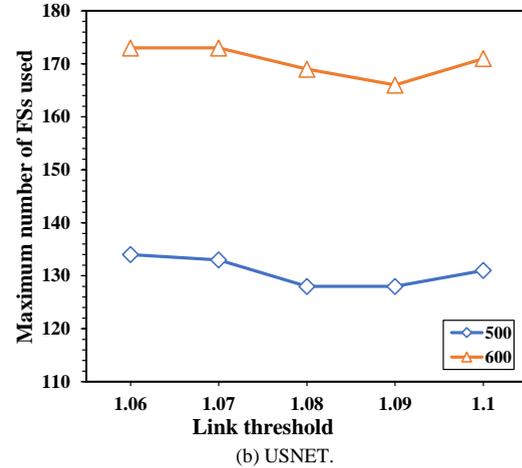

(b) USNET.
FIGURE 7. Maximum number of FSs used under different link thresholds α.

### D. SPECTRUM AND COST EFFICIENCY OF HUS-EON

Deploying ULL fibers on network links can enhance network performance, but it also significantly increases the network cost due to high expense of ULL fibers. This section evaluates whether the HUS-EON can well balance network performance and fiber deployment cost. We consider four different network scenarios, including only one SSMF in each link, two SSMFs in each link, two ULL fibers in each link, and one ULL and one SSMF in each link (i.e., HUS-EON). Fig. 8 presents the lightpath blocking probabilities and costs for the USNET network under different network scenarios, where the traffic load between each node pair ranges from 3.2 Erlang to 4.4 Erlang. The legends "S" represents the network with only one SSMF in each link, "SS" represents the network with two SSMFs in each link, "US" represents to network with one ULL fiber and one SSMF in each link, "UU" represents network with two ULL fibers in each link.

We can observe that increasing the number of fibers on each link can significantly reduce the lightpath blocking probability. This is reasonable that more fibers provide more spectrum resources and reduce the lightpath blocking probability. In addition, we also see that the more ULL fibers there are on each link, the lower the lightpath blocking probability.

Specifically, compared to the network with only one SSMF in each link, the networks with two ULL fibers, one ULL fiber and one SSMF, and two SSMFs in each link can achieve lightpath blocking performance improvements of 98.8%, 94.5%, and 81.2% respectively. This is due to the fact that more ULL fibers enable more traffic demands to require fewer FSs, thereby enhancing spectrum utilization efficiency and reducing the lightpath blocking probability.

Meanwhile, we also evaluate the total fiber cost of different network scenarios, calculated as the $\sum_l D_l \cdot (n_l^s \cdot \varsigma_s + n_l^u \cdot \varsigma_u)$. $D_l$ is the physical distance of link $l$ in kilometer. $n_l^s$ and $n_l^u$ are respectively the number of newly deployed SSMFs and ULL fibers in link $l$. $\varsigma_s$ and $\varsigma_u$ are the costs per kilometer of SSMF and ULL fiber respectively. Note that $\varsigma_s$ and $\varsigma_u$ are assumed to be normalized to 1 units/km and 10 units/km, respectively. The total fiber deployment cost of the three network scenarios are 8584 units, 85840 units, 171680 units respectively. We find that deploying one ULL and one SSMF on the network link (i.e., HUS-EON) is more cost-effective than deploying two ULL fibers in each link, since it can achieve nearly the same lightpath blocking probability performance at half the total fiber cost.

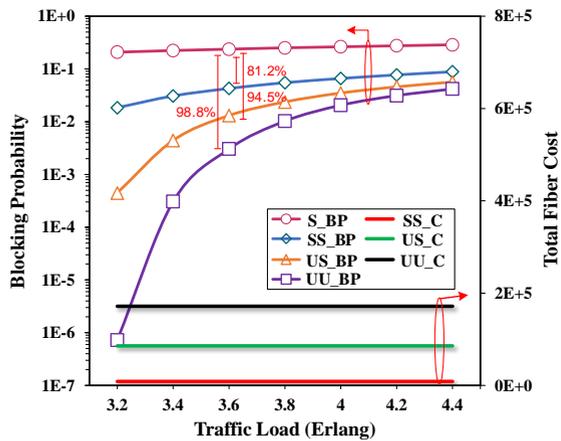

**FIGURE 8.** Blocking probability and total fiber cost of different network scenarios.

## VII. CONCLUSION

In this study, we addressed the RFMSA problem in an EON context where ULL fiber and SSMF coexist on each link. We formulated the RFMSA problem as a node-arc based MILP model and developed efficient SWP-based algorithms to solve the problem. For both static and dynamic traffic demand scenarios, we considered different strategies for selecting proper fibers to establish lightpaths, including Random, UFF, OA, and SU strategies respectively. Note that the OA strategy is only suitable for static traffic demand, and the SU strategy is exclusively used for dynamic traffic demand. Simulation results indicate that in the static traffic demand scenario, the SWP-based RFMSA algorithm with OA strategy performs very similarly to the MILP model and consistently outperforms the Random and UFF strategies. In the dynamic traffic demand scenarios, the SU strategy proves to be superior to the Random and UFF strategies, demonstrating its effectiveness. Moreover, the performance of the SWP-based RFMSA algorithm is significantly better than that of the SP-based RFMSA algorithm, highlighting substantial improvements in spectrum efficiency. Finally, by comparing the lightpath blocking probability and total fiber cost, we note that the HUS-EON is more cost-efficient compared to the EONs with all SSMFs or ULL fibers, respectively.

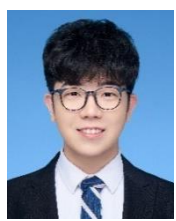

**Kangao Ouyang** received his B.S. degree in communication engineering from Soochow University, Suzhou, China, in 2022. He has been studying in the School of Electronics and Information Technology, Soochow University, Suzhou, China, since 2022. His research interests include elastic optical networks, and network design and optimization.

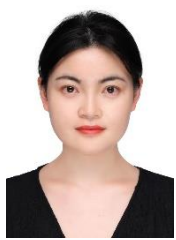

**Fengxian Tang** received her PhD degree at Soochow University, in 2021. From 2019 to 2020, she was a visiting scholar at North Carolina State University. Now, she is an associated professor at Shenzhen Polytechnic University. Her research interest includes SDM optical network, Data center network.

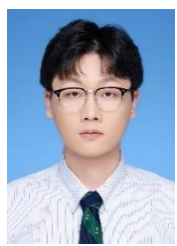

**Zhilin Yuan** received the B.S. degree in electronic information science and technology from Nanjing Institute of Technology, Nanjing, China, in 2022. He has been studying in the School of Electronics and Information Technology, Soochow University, Suzhou, China, since 2022. His research interests include optical switching architectures, and network design and optimization.

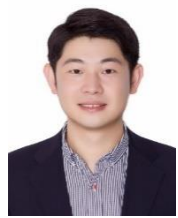

**Jun Li** received his PhD degree at KTH Royal Institute of Technology, in 2019. From 2018 to 2019, he was a visiting scholar at Princeton University. Now, he is an associated professor at Soochow University. His research interests include optical access network, 6G optical transport networks. He has received 2017 IEEE/Optical ACP best student paper award and 2023 IEEE/Optical OECC Best paper award.

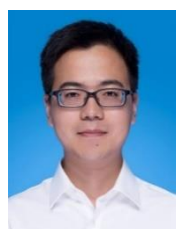

**Yongcheng Li** received his B.Sc. degree and Ph.D. from Soochow University, China, in 2011 and 2017, respectively. Currently, he is an associated researcher with the School of Electronic and Information Engineering of Soochow University. His research interests include optical switching, optical networking, and network design and optimization. He has received 2013 IEEE/Optical ACP best student paper award.